\documentclass[twoside,twocolumn,9pt]{article}
\usepackage{extsizes}
\usepackage[super,sort&compress,comma]{natbib} 
\usepackage[version=3]{mhchem}
\usepackage[left=1.5cm, right=1.5cm, top=1.785cm, bottom=2.0cm]{geometry}
\usepackage{balance}
\usepackage{times,mathptmx}
\usepackage{sectsty}
\usepackage{graphicx} 
\usepackage{lastpage}
\usepackage[format=plain,justification=justified,singlelinecheck=false,font={stretch=1.125,small,sf},labelfont=bf,labelsep=space]{caption}
\usepackage{float}
\usepackage{fancyhdr}
\usepackage{fnpos}
\usepackage[english]{babel}
\addto{\captionsenglish}{%
  
}
\usepackage{array}
\usepackage{droidsans}
\usepackage{charter}
\usepackage[T1]{fontenc}
\usepackage[usenames,dvipsnames]{xcolor}
\usepackage{setspace}
\usepackage[compact]{titlesec}
\usepackage{hyperref}
\definecolor{cream}{RGB}{222,217,201}

\begin{document}

%%%PAGE SETUP - Please do not change any commands within this section%%%
\makeFNbottom
\makeatletter
\renewcommand\LARGE{\@setfontsize\LARGE{15pt}{17}}
\renewcommand\Large{\@setfontsize\Large{12pt}{14}}
\renewcommand\large{\@setfontsize\large{10pt}{12}}
\renewcommand\footnotesize{\@setfontsize\footnotesize{7pt}{10}}
\makeatother

\renewcommand{\thefootnote}{\fnsymbol{footnote}}
\renewcommand\footnoterule{\vspace*{1pt}% 
\color{cream}\hrule width 3.5in height 0.4pt \color{black}\vspace*{5pt}} 
\setcounter{secnumdepth}{5}

\makeatletter 
\renewcommand\@biblabel[1]{#1}            
\renewcommand\@makefntext[1]% 
{\noindent\makebox[0pt][r]{\@thefnmark\,}#1}
\makeatother 
\renewcommand{\figurename}{\small{Fig.}~}
\sectionfont{\sffamily\Large}
\subsectionfont{\normalsize}
\subsubsectionfont{\bf}
\setstretch{1.125} %In particular, please do not alter this line.
\setlength{\skip\footins}{0.8cm}
\setlength{\footnotesep}{0.25cm}
\setlength{\jot}{10pt}
\titlespacing*{\section}{0pt}{4pt}{4pt}
\titlespacing*{\subsection}{0pt}{15pt}{1pt}
%%%END OF PAGE SETUP%%%

%%%FOOTER%%%
\fancyfoot{}
\fancyfoot[LO,RE]{\vspace{-7.1pt}\includegraphics[height=9pt]{head_foot/LF}}
\fancyfoot[CO]{\vspace{-7.1pt}\hspace{13.2cm}\includegraphics{head_foot/RF}}
\fancyfoot[CE]{\vspace{-7.2pt}\hspace{-14.2cm}\includegraphics{head_foot/RF}}
\fancyfoot[RO]{\footnotesize{\sffamily{1--\pageref{LastPage} ~\textbar  \hspace{2pt}\thepage}}}
\fancyfoot[LE]{\footnotesize{\sffamily{\thepage~\textbar\hspace{3.45cm} 1--\pageref{LastPage}}}}
\fancyhead{}
\renewcommand{\headrulewidth}{0pt} 
\renewcommand{\footrulewidth}{0pt}
\setlength{\arrayrulewidth}{1pt}
\setlength{\columnsep}{6.5mm}
\setlength\bibsep{1pt}
%%%END OF FOOTER%%%

%%%FIGURE SETUP - please do not change any commands within this section%%%
\makeatletter 
\newlength{\figrulesep} 
\setlength{\figrulesep}{0.5\textfloatsep} 

\newcommand{\topfigrule}{\vspace*{-1pt}% 
\noindent{\color{cream}\rule[-\figrulesep]{\columnwidth}{1.5pt}} }

\newcommand{\botfigrule}{\vspace*{-2pt}% 
\noindent{\color{cream}\rule[\figrulesep]{\columnwidth}{1.5pt}} }

\newcommand{\dblfigrule}{\vspace*{-1pt}% 
\noindent{\color{cream}\rule[-\figrulesep]{\textwidth}{1.5pt}} }

\makeatother
%%%END OF FIGURE SETUP%%%

%%%TITLE, AUTHORS AND ABSTRACT%%%
\twocolumn[
  \begin{@twocolumnfalse}
\sffamily
\vspace{2cm}
\noindent\LARGE{\textbf{Structural phase transitions in VSe$_2$: energetics, electronic structure and magnetism$^\dag$}} \\%Article title goes here instead of the text "This is the title"

 \noindent\large{Georgy V. Pushkarev,\textit{$^{a}$} Vladimir G. Mazurenko,\textit{$^{a}$} Vladimir V. Mazurenko\textit{$^{a}$} and Danil W. Boukhvalov $^{\ast}$\textit{$^{a,b}$}} \\%Author names go here instead of "Full name", etc.

\vspace{2cm}

 \noindent\normalsize{First principles calculations of magnetic and electronic properties of VSe$_2$ describing the transition between two structural phases (H,T) were performed. Results of the calculations  evidence rather low energy barrier (~0.60 eV for monolayer) for transition between the phases. The energy required for the deviation of Se atom or whole layer of selenium atoms on a small angle up to 10$^\circ$ from initial positions is also rather low, 0.32 and 0.19 eV/Se, respectively. The changes in band structure of VSe$_2$ caused by these motions of Se atoms should be taken into account for analysis of the experimental data. Simulations of the strain effects suggest that the experimentally observed T phase of VSe$_2$ monolayer is the ground state due a substrate-induced strain. Calculations of the difference in total energies of ferromagnetic and antiferromagnetic configurations evidence that the ferromagnetic configuration is the ground state of the system for all stable and intermediate atomic structures. Calculated phonon dispersions suggest visible influence of magnetic configurations on vibrational properties.} 

 \end{@twocolumnfalse} \vspace{1cm}

  ]
%%%END OF TITLE, AUTHORS AND ABSTRACT%%%

%%%FONT SETUP - please do not change any commands within this section
\renewcommand*\rmdefault{bch}\normalfont\upshape
\rmfamily
\section*{}
\vspace{-1cm}

%%%FOOTNOTES%%%

\footnotetext{\textit{$^{a}$~Ural Federal University
620002, 19 Mira street, Ekaterinburg, Russia. Tel: +7 343 375 4444; E-mail: contact@urfu.ru}}
\footnotetext{\textit{$^{b}$~College of Science, Institute of Materials Physics and Chemistry, Nanjing Forestry University, Nanjing 210037, PR China }}

%Please use \dag to cite the ESI in the main text of the article.
%If you article does not have ESI please remove the the \dag symbol from the title and the footnotetext below.
%\footnotetext{\dag~Electronic Supplementary Information (ESI) available: [details of any supplementary information available should be included here]. See DOI: 00.0000/00000000.}
%additional addresses can be cited as above using the lower-case letters, c, d, e... If all authors are from the same address, no letter is required

%\footnotetext{\ddag~Additional footnotes to the title and authors can be included \textit{e.g.}\ `Present address:' or `These authors contributed equally to this work' as above using the symbols: \ddag, \textsection, and \P. Please place the appropriate symbol next to the author's name and include a \texttt{\textbackslash footnotetext} entry in the the correct place in the list.}

%%%END OF FOOTNOTES%%%

%%%MAIN TEXT%%%%
\section{INTRODUCTION}
Monolayer VSe$_2$ is the one of the most intriguing members of the family of two-dimensional (2D) transition-metal dichalcogenides. This material attracts a special interest of the scientific community due to several recent discoveries, including in-plane piezoelectricity \cite{1}, a pseudogap with Fermi arc \cite{2} at temperatures above the charge density wave transition (~220 K for the monolayer \cite{3}), and especially the existence of ferromagnetism in 2D system\cite{4,5,6,7,8,9,10,11}. Experimental results are rather contradictory. A strong room-temperature ferromagnetism with a huge magnetic moment per formula unit has been reported for monolayer VSe$_2$ epitaxially grown on graphite \cite{4}. A local magnetic phase contrast has also been observed by magnetic force microscopy at the room temperature at the edges of VSe$_2$ flakes exfoliated from a three-dimensional crystal. \cite{14} XMCD measurements evidence a spin-frustrated magnetic structure in VSe$_2$ on graphite. \cite{xmcd} Paramagnetism of bulk VSe$_2$ \cite{para1,para2} makes these observations more intriguing. Another situation was reported for the monolayers grown on bilayer graphene/silicon carbide substrate. In both works the absence of exchange splitting of the vanadium $3d$ bands observed in angle-resolved photoemission spectroscopy experiments was reported. This result contradicts to other studies that revealed a magnetization value not higher than ~5 $\mu_B$.\cite{12,13} Based on these results we can conclude that the influence of the substrate is important for description of the magnetic properties of these materials. Theoretical models have been developed to account for the above discrepant observations \cite{4,12,14,16}. These works mainly focused on the band structure and magnetic moments on vanadium sites. It has been proposed that the presence of charge density waves could cause the quenching of monolayer ferromagnetism due to the band gap opening induced by Peierls distortion \cite{15}. Phonon spectra of several VSe$_2$ and similar systems
were also considered theoretically \cite{22,23}. This modeling motivates us to study interplay between magnetism and structural phase transitions in VSe$_2$. Additionally, there is a plethora of works demonstrating a relationship between the symmetry, electronic structure and magnetic properties in transitional metal compounds \cite{BaCu, DMI1, DMI2,24}.

\begin{figure}[t]
    \centering
    \includegraphics[width=1\columnwidth]{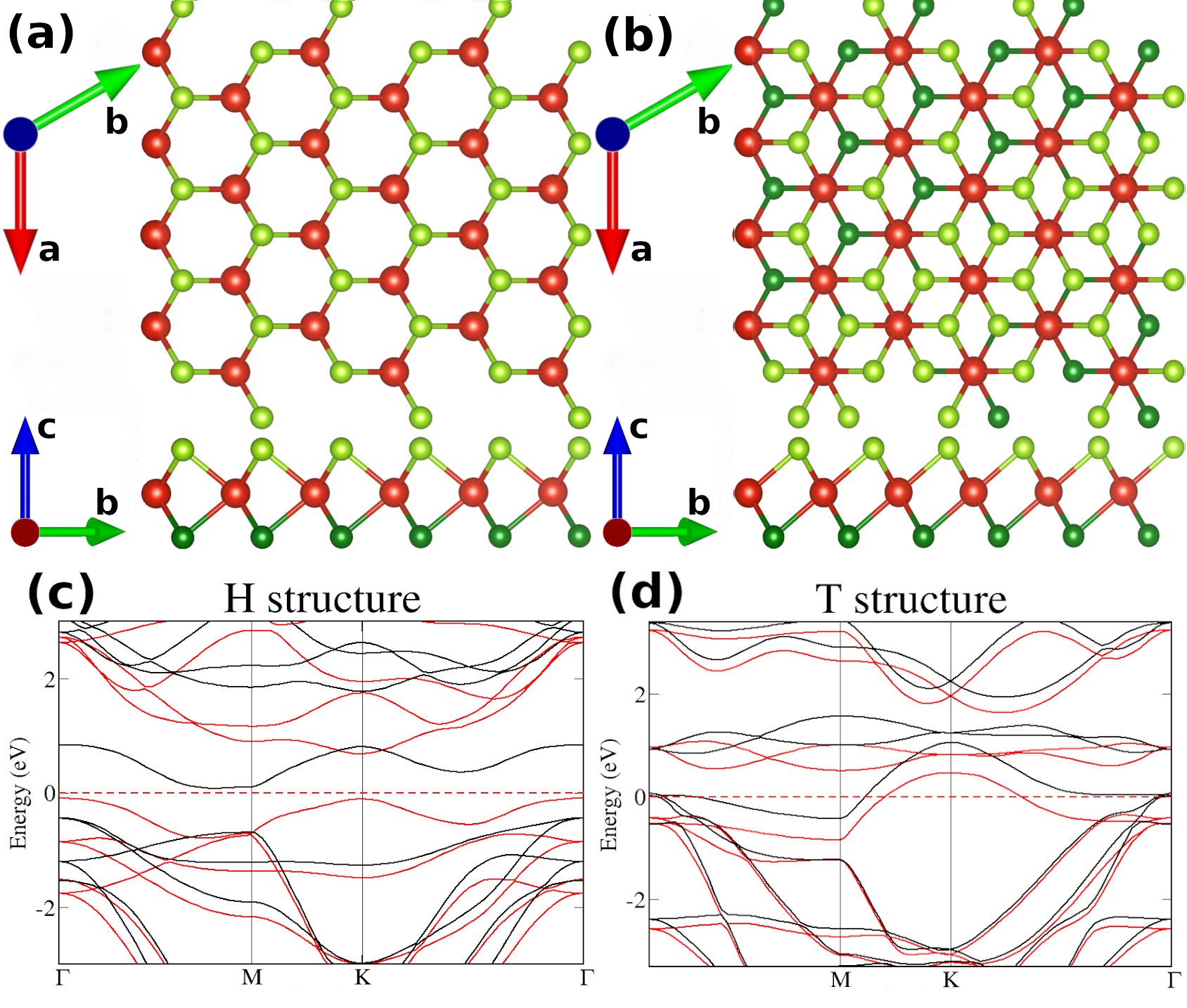}
    \caption{Atomic structure of 2D VSe$_2$ monolayer (top and side view) in H phase (a) and in T phase (b). Vanadium atoms are denoted with red circles, upper and bottom selenium layers denoted with light green and dark green circles, respectively. (c) and (d) panels represent the corresponding spin-polarized band structures. Red lines correspond to spin up states and black ones to spin down, the Fermi level corresponds to 0 eV.}
    \label{fig1}
\end{figure}

The VSe$_2$ crystal is formed from separate layers along the c-axis direction. Two main phases for this material were predicted to be stable: the H phase characterized by Se stacked over each other and the T phase with one layers of Se rotated by $60^\circ$ around axis normal to the plane of layer. \cite{16} Atomic structures of the VSe$_2$ monolayer in both H and T phases are shown in Fig.1. Surprisingly, the reported binding energies for different configurations are almost the same despite the colossal difference in magnetic properties and electronic structure (Fig.\ref{fig1}). \cite{16} This finding additionally motivates us to examine various aspects of structural phase transitions in bulk, few-layer and monolayer of VSe$_2$.
\newpage
%%%%%%%%%%%%%%%%%%%%%%%%%%%%%%%%%%%%%%%%%%%%%%%%%%%%%%%%%%%%%%%%%%%%%%%%%%%%%%%%%%%%%%%%%%%%%%
%%%%%%%%%%%%%%%%%%%%%%%%%%%%%%%%%%%%%%%%%%%%%%%%%%%%%%%%%%%%%%%%%%%%%%%%%%%%%%%%%%%%%%%%%%%%%%
%%%%%%%%%%%%%%%%%%%%%%%%%%%%%%%%%%%%%%%%%%%%%%%%%%%%%%%%%%%%%%%%%%%%%%%%%%%%%%%%%%%%%%%%%%%%%%
\section{Computational method and model}
Electronic properties of the VSe$_2$ system were
simulated within Density Functional Theory (DFT) framework using the Perdew-Burke-Ernzerhof (PBE) exchange-correlation functional \cite{17} as implemented in the Vienna ab-initio simulation
package (VASP) \cite{19,20} with a plane-wave basis set. This approach gave reliable results for other systems similar to VSe$_2$ \cite{25}. Also we include van der Waals interaction using the method of Grimme (DFT (PBE)-D2) \cite{vdW}. Taking into account London
dispersion forces is essential for few-layer VSe$_2$ (see Table
\ref{tab1} and discussion in section 3.5). 

The calculation parameters were chosen as follows. The energy cutoff
equals to 400 eV and the energy convergence criteria is
$10^{-6}$ eV. For the Brillouin zone integration a $10\times10\times1$
gamma centered grid was used for layered structures and $8\times8\times8$ for bulk structures. A vacuum space more than 10 \AA \, in the vertical $z$ direction
was introduced for layered structures. The technical parameters are similar to those used in the recent studies of phase stability in layered systems.\cite{Ersan, Kaltsas}

The optimized atomic positions for T-phase and lattice parameters $a=b=3.31$ \AA \, and $c=6.20$ \AA \, are in good agreement with experiment \cite{21}.
In particular, the corresponding interlayer distance in bulk VSe$_2$ is $3.04$ \AA. The calculated band structures of VSe$_2$  monolayer in the T and H phases are in good agreement with previous works. \cite{16} The calculated magnetic moment of 0.68 $\mu_B$ for initial configuration without rotation of the selenium atoms also agrees with results of the previous work \cite{18}.

\begin{figure}[h]
\includegraphics[width=0.875\columnwidth]{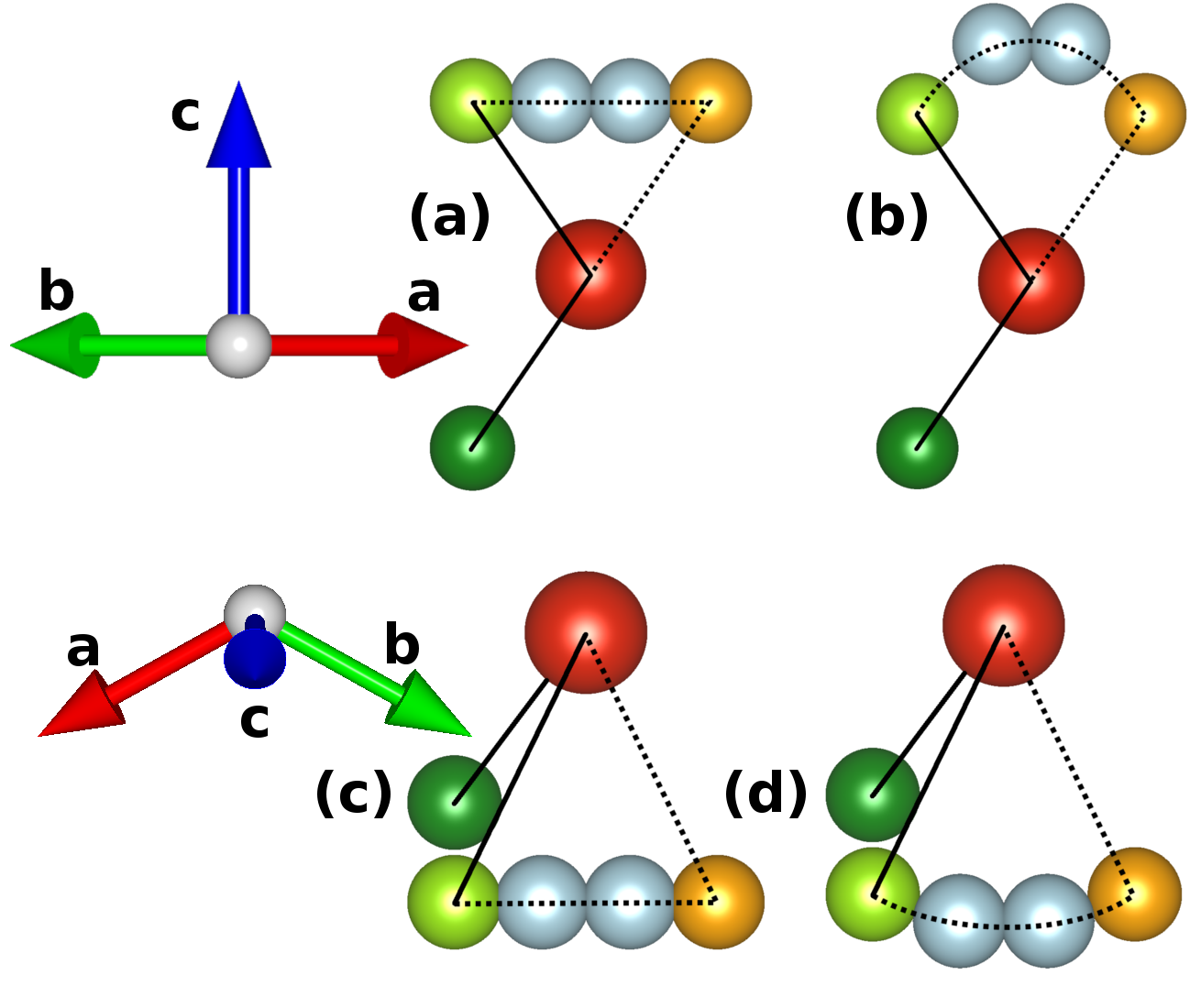}
\caption{Schematic visualization of the plane (a,c) and arc (b,d) types of the Se atoms rotation. (a,b) and (c,d) panels correspond to side and top views, respectively. Initial and final positions of Se are presented with orange and green circles, respectively. Intermediate configurations of selenium atoms obtained with the 20$^{\circ}$ step are denoted with light blue circles.}
    \label{fig2}
\end{figure}

To investigate the transition between H and T phases we performed self-consistent calculations of electronic structure and total energies in transitional points between these phases. For this purpose, we rotate either one Se atom or all selenium atoms belonging to the upper layer of VSe$_2$ in supercell as schematically shown in Fig.\ref{fig2}. To trace the changes in electronic structure and magnetic properties the calculations for configurations with a 10$^{\circ}$ rotation step were performed. Generally, the rotation can be realized within two models. The first one is to move Se in plane from initial to final point (Fig.\ref{fig2} a and c). The second one is to fix the constant V-Se distance for all intermediate steps, which produces an elevation of selenium atoms above the plane at intermediate steps of the migration (Fig.\ref{fig2} b and d). We will refer these rotation models as in-plane and arc rotation schemes, respectively. All the calculations were performed for the ferromagnetic ordering of the spins of vanadium atoms. 

%%%%%%%%%%%%%%%%%%%%%%%%%%%%%%%%%%%%%%%%%%%%%%%%%%%%%%%%%%%%%%%%%%%%%%%%%%%%%%%%%%%%%%%%%%%%%%
%%%%%%%%%%%%%%%%%%%%%%%%%%%%%%%%%%%%%%%%%%%%%%%%%%%%%%%%%%%%%%%%%%%%%%%%%%%%%%%%%%%%%%%%%%%%%%
%%%%%%%%%%%%%%%%%%%%%%%%%%%%%%%%%%%%%%%%%%%%%%%%%%%%%%%%%%%%%%%%%%%%%%%%%%%%%%%%%%%%%%%%%%%%%%

\section{Results and discussion}
\subsection{Rotation of single Se atom}
\begin{figure}[ht]
    \centering
    \includegraphics[width=1\columnwidth]{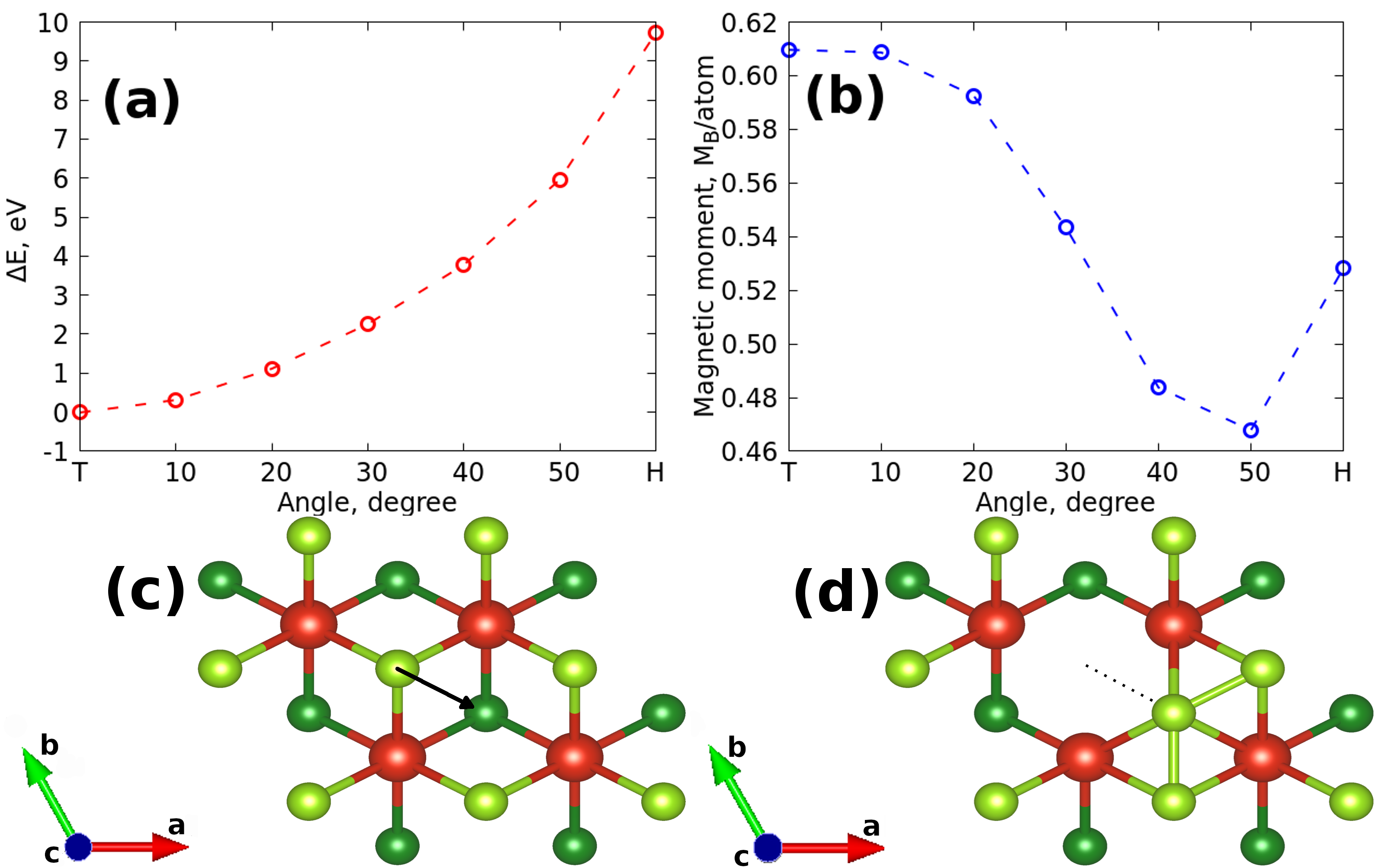}
    \caption{Evolution of the total energy (a) and magnetic moment (b) during in-plane rotation of single Se atom. (c) and (d) panels visualize the initial and final atomic structures. Light and dark green circles denote upper and bottom selenium layers, respectively.}
    \label{fig3}
\end{figure}

At the first step of our study we have simulated the motion of the single Se atom in the monolayer (see Fig. \ref{fig3}). For simplicity, we considered an in-plane migration of the atom. Results of the calculations (Fig. 3) evidence a gradual increasing of the total energy of the system during the all processes of the rotation with maximal value at final point. The cause of the large magnitude of the energies and instability of the final configuration is in decreasing of the distance between moved and rigid Se atoms to the value of 1.92 \AA. Thus we can conclude that the model of the single Se atom rotation is unrealistic and transition between T and H phases may be realized only with distortion of the whole selenium layer. Further we will consider only this kind of the structural phase transitions. The values of the magnetic moments calculated for intermediate configurations (Fig. \ref{fig3}) support our initial guess that the structural transition between the phases affects magnetic properties of VSe$_2$. Note that a deviation of the selenium atoms from equilibrium positions on small angles (less than 10$^{\circ}$) requires much smaller energies of about 0.32 eV and, therefore, should be taken into account for a realistic description of the atomic structure of VSe$_2$ at the room temperature.

%%%%%%%%%%%%%%%%%%%%%%%%%%%%%%%%%%%%%%%%%%%%%%%%%%%%%%%%%%%%%%%%%%%%%%%%%%%%%%%%%%%%%%%%%%%%%%
%%%%%%%%%%%%%%%%%%%%%%%%%%%%%%%%%%%%%%%%%%%%%%%%%%%%%%%%%%%%%%%%%%%%%%%%%%%%%%%%%%%%%%%%%%%%%%
%%%%%%%%%%%%%%%%%%%%%%%%%%%%%%%%%%%%%%%%%%%%%%%%%%%%%%%%%%%%%%%%%%%%%%%%%%%%%%%%%%%%%%%%%%%%%%
\subsection{Rotation of the whole Se sheet in the VSe$_2$ monolayer}
 Having considered the results concerning the migration of the single Se atom we are in a position to analyze the case of whole upper Se-layer rotation, which will provide a better understanding of the transition between H and T phases.
 \begin{figure}[ht]
    \centering
    \includegraphics[width=1\columnwidth]{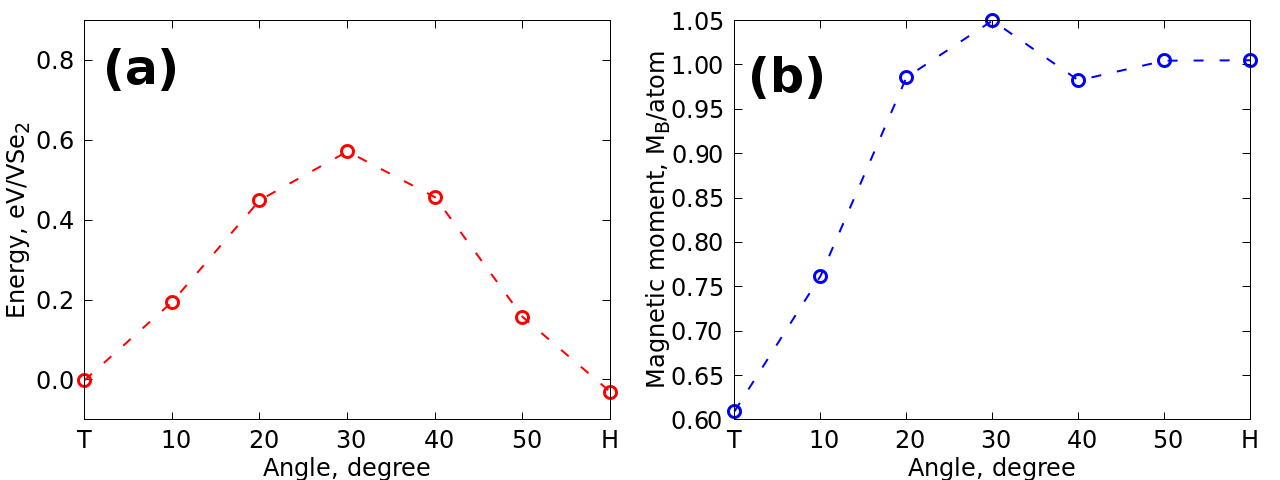}
    \caption{Evolution of the total energy (a) and magnetic moment (b) during rotation of whole upper Se layer of VSe$_2$ monolayer within arc model.}
    \label{fig4}
\end{figure}

 The performed simulations for 3$\times$3 supercell with constant V-Se distances when Se atoms elevate from initial and starting positions (Fig. \ref{fig2}) revealed the energy barrier of 0.60 eV that is smaller than that observed in the case of the in-plane rotation (Fig.\ref{figS1}a in SI). Thus further we will consider only this type of the Se atom migration. To evaluate the temperature required to overcome this barrier one should establish a relation between the calculated energies of the process and temperature of the reactions. We have addressed this question in our previous work \cite{go} and found that the barrier values of about 0.50 eV and 1.20 eV correspond to the room temperature and 200 $^{\circ}$C, respectively. Thus, the energy barrier of 0.60 eV can be overcome already at the temperatures about 40 $^\circ$C. 
 
 Four conclusions could be drawn from these results. (i) There is a possibility of the structural phase transition in previously studied VSe$_2$ samples during measurements. (ii) For development of devices based on VSe$_2$ and similar monolayer systems one should take into account possibility of the structural phase transitions caused by the heating of the devices during work. Such a transition can significantly affect the work of the device due to difference in electronic structures of different  phases (see Fig. \ref{fig1},also changes in band structure Fig.\ref{figS3} in SI). (iii) One can use VSe$_2$ and similar systems as temperature detectors. (iv) According to our results there is a low-energy cost to deviate the selenium atoms belonging to one layer on a small angle from the equilibrium positions. It means that one needs to account this for a realistic interpretation of the experimental data.

\begin{figure}[ht]
    \centering
    \includegraphics[width=1\columnwidth]{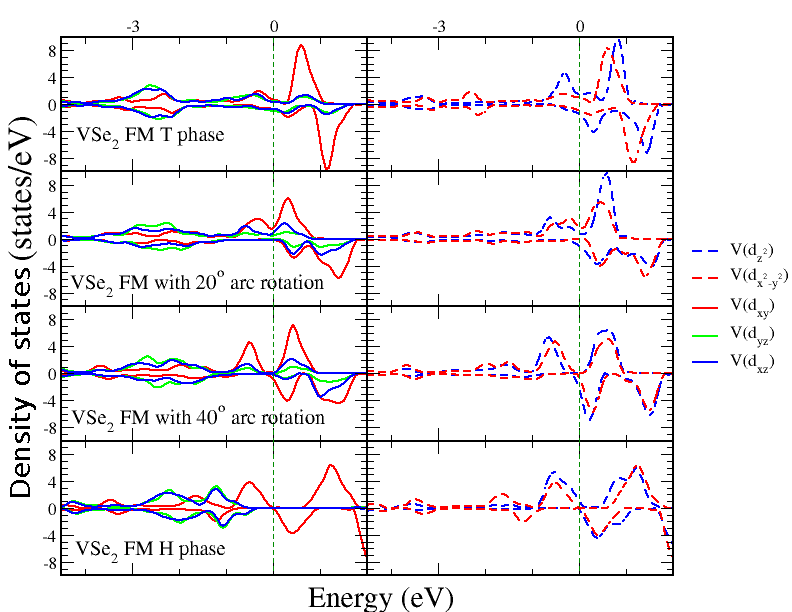}
    \caption{Partial densities of states calculated for VSe$_2$ monolayer in the ferromagnetic configuration. The arc rotation scheme  with the $20^{\circ}$ step was used. Left and right panels correspond to ($d_{xy}$, $d_{yz}$, $d_{xz}$) and ($d_{x^2-y^2}$, $d_{z^2}$) sets of states, respectively.   }
    \label{fig5}
\end{figure}

Moderate temperature of the transition between different structural phases requires an examination of the electronic structure and magnetic properties at intermediate steps of the structural phase transition. The obtained calculations results demonstrate that in the case of the ferromagnetic ground state the values of magnetic moments change gradually with small step of 10$^\circ$ of the rotation of Se layer. From Fig.\ref{fig4} one can see that at 30$^\circ$ the magnetic moment has the maximal value of 1.05$\mu_{B}$, which is about two times larger than that in the initial configuration. According to the calculated occupation matrices such a magnetic moment change is mainly related to the contributions of $xy$ and $x^2-y^2$ orbitals of vanadium atoms (see Fig. \ref{fig5}). Since the total occupation (spin-up + spin-down) of the different orbitals remain almost the same, the orbital magnetic moment values change is fully connected with a redistribution of the electrons between different spin channels due to change of the hybridization between V and Se. 

In the case of an antiferromagnetic configuration the situation is more complicated. First of all, the magnetic lattice of VSe$_2$ is frustrated one, which is in agreement with experimental observations.\cite{xmcd}  This means that within a mean-field DFT approach we cannot define an antiferromagnetic collinear-type order corresponding to minimum of the magnetic interaction energy for all V-V bonds, simultaneously. The second complication follows from the fact that the system in question is metal. It means that the magnetization of individual vanadium atom can be very sensitive to the orientation of the neighbouring magnetic moments \cite{FeGe}. Indeed, our DFT simulations of the VSe$_2$ supercell with antiferromagnetic ordering have revealed a strong suppression of the magnetic moment values of some vanadium atoms in the supercell. In addition, we observe that the details of the magnetic moments suppression strongly depend on the size of the supercell. In this complex situation some information on magnetic couplings in the VSe$_2$ system could be extracted by using the theory of infinitesimal spin rotations approximation \cite{FeGe, Liechtenstein}. However, the magnetic couplings calculated in this way can be used for analysis only in the vicinity of the ferromagnetic configuration. 

The values of the magnetic moments in AFM phase can be stabilized by inclusion of the on-site Coulomb interaction as can be done with DFT+$U$ approach. However, the using of the DFT+$U$ approach in the case of VSe$_2$ is questionable, since the experimental ARPES spectra are in good agreement with GGA band structure as it was shown in Refs.\cite{xmcd,arp1,12}. At the same time the inclusion of the Hubbard $U$ leads to considerable changes in the band structure.

\begin{figure}[ht]
    \centering
    \includegraphics[width=0.92\columnwidth]{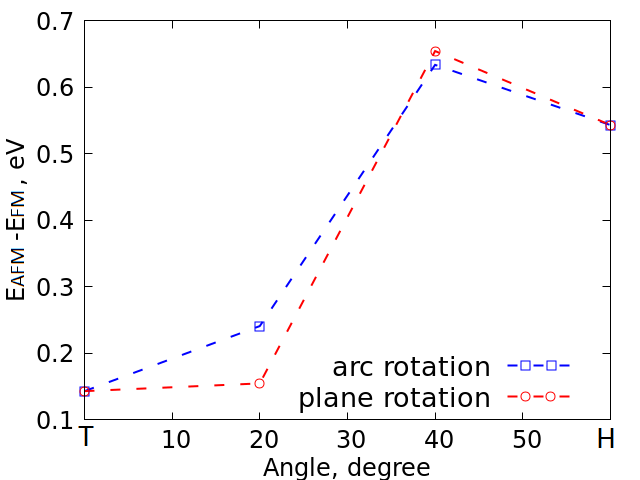}
    \caption{Difference of AFM and FM state total energies calculated in arc and plane rotation schemes for 3x3 supercell of  VSe$_2$ monolayer.}
    \label{fig6}
\end{figure}

Thus, the energy difference between AFM and FM solutions for VSe$_2$ simulated with GGA does not allow us to construct a comprehensive magnetic model and estimate the corresponding magnetic interactions between vanadium atoms. Nevertheless, the results of these calculations evidence that despite the changes of electronic structure at intermediate steps the ferromagnetic configuration remains significantly energetically favorable in all the cases (Fig.\ref{fig6}). Thus the possible structural distortions in VSe$_2$ will not provide a suppression of ferromagnetism. Our calculations demonstrate that possible transition from experimentally observed T phase toward H phase should provide an enhancement of ferromagnetic interactions and increasing of magnetic moment. To simulate the experimentally observed paramagnetic state of bulk VSe$_2$ \cite{para1,para2} one can use a dynamical mean-field theory.

%%%%%%%%%%%%%%%%%%%%%%%%%%%%%%%%%%%%%%%%%%%%%%%%%%%%%%%%%%%%%%%%%%%%%%%%%
%%%%%%%%%%%%%%%%%%%%%%%%%%%%%%%%%%%%%%%%%%%%%%%%%%%%%%%%%%%%%%%%%%%%%%%%%
\subsection{Structural phase transition in bulk VSe$_2$}
\begin{figure}[ht]
    \includegraphics[width=1\columnwidth]{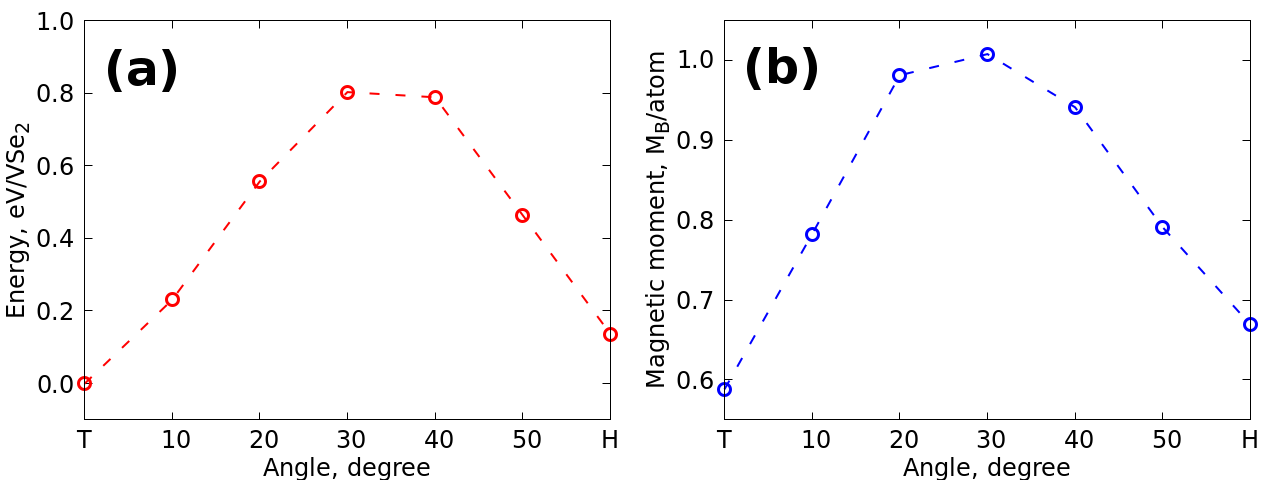}
    \caption{Total energy (a) and magnetic moment (b) as functions of the rotation angles. The simulation results were obtained for bulk VSe$_2$ within the arc rotation model.}
    \label{fig7}
\end{figure}

There are two main differences in the energetics of the structural phases transitions in bulk and monolayer VSe$_2$. The first one is the almost the same value for the energies of the motion of Se layer within both rotation models (Fig.\ref{fig7} and Fig.\ref{figS1}c in SI). The second one is increasing of the migration barrier (see Fig. \ref{fig7}). Both are related to the van der Waals interactions between the layers in the bulk VSe$_2$. The analysis of the calculated partial density of states in this case leads to similar conclusions as above (see Fig.\ref{figS2} in SI)

In the case of the rotation of the Se atoms belonging to one layer with constant V-Se distance at intermediate steps of migration, the initial distance of 3.63 \AA \, between rotated and fixed selenium layers decreases by 0.54 \AA. This deviation from the optimal interlayer distance provides an increasing of the energy barrier (see also changes in band structure Fig.\ref{figS4} in SI).  The value of the energy barrier is corresponding with stability of the structural ground state in bulk crystal up to the temperatures above 100$^{\circ}$C. Note that in contrast to monolayer case the structural ground state of bulk VSe$_2$ is T configuration with  ferromagnetic orientation of magnetic moments.

\subsection{Structural phase transition in bi- and trilayers of VSe$_2$}
\begin{figure}[ht]
    \includegraphics[width=1\columnwidth]{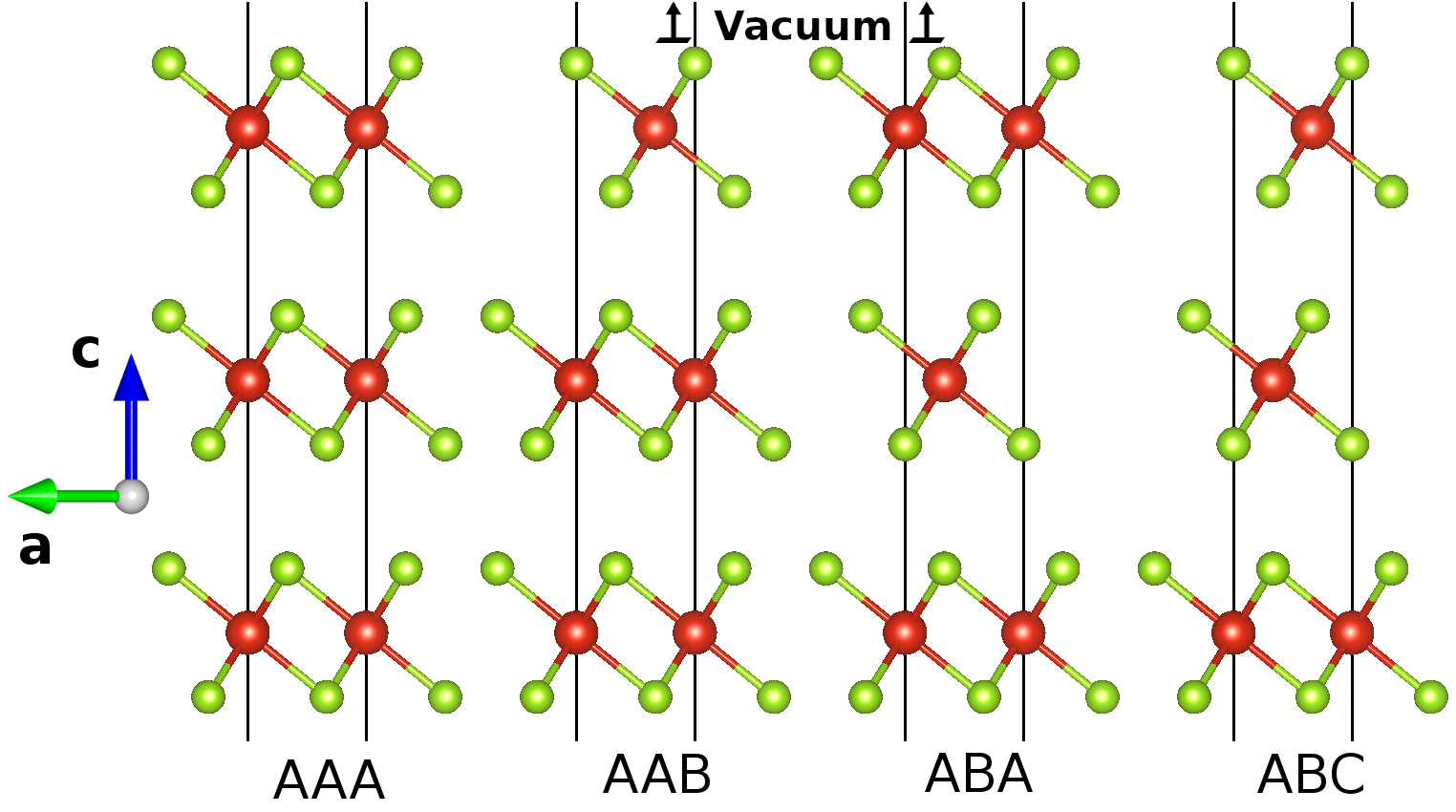}
    \caption{Schematic representation of the unit cells used for simulating VSe$_2$ trilayers characterized by different stacking models.}
    \label{fig8}
\end{figure}

Moreover, we examine the energetics of the structural phases transition in the top layer of bi- and trilayers VSe$_2$ with different stacking models (Fig. \ref{fig8}). The notation of the types of Bernal stacking is similar to graphite. These results also can be applied for VSe$_2$ non-covalently attached to substrates.

\begin{figure}[ht]
    \centering
    \includegraphics[width=1\columnwidth]{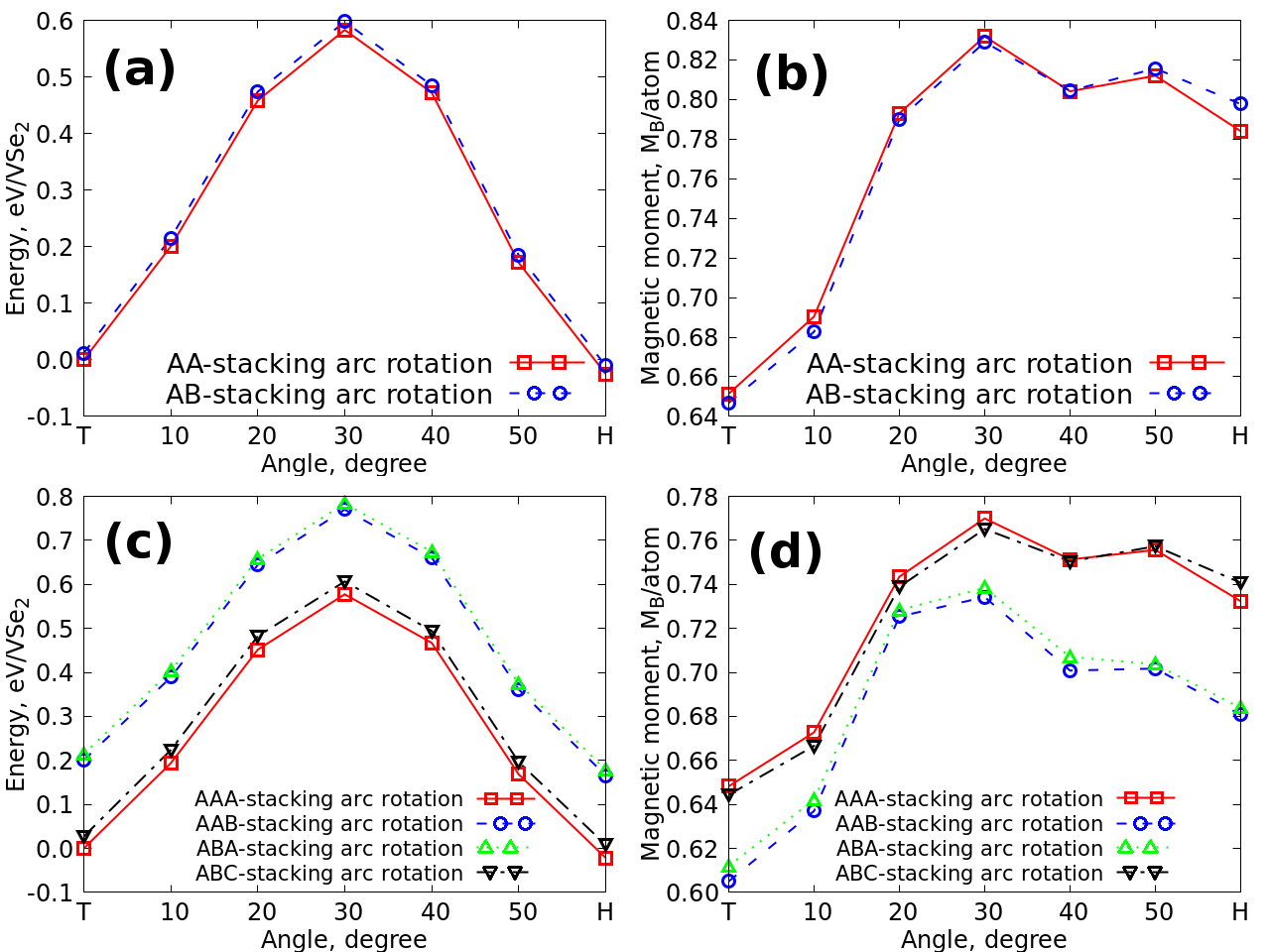}
    \caption{Total energy (left panels) and magnetic moment (right panels) of two- (a,b) and three-layer (c,d) VSe$_2$ systems estimated for H, T and intermediate structures.}
    \label{fig9}
\end{figure}

Results of the calculations (Fig. \ref{fig9}) evidence similarity the case of few-layer VSe$_2$ with monolayer. The configuration of the H type corresponds to the structural ground state for all types of the stacking in few-layer case. The energy required for the transition from T to H phase is about 0.60 eV for AA- and AB- stacking in bilayer. In trilayer the most energetically favorable stacking orders are AAA and ABC.

 Thus, similarly to free standing VSe$_2$ monolayer, in top layer of VSe$_2$ there can be transition between two structural configurations at a moderate heating. The magnetic moments of vanadium atoms belonging to the upper layer of the few-layer structures change from 0.64$\mu_B$ to 0.82$\mu_B$. Such a change is fully connected with a redistribution of the electrons between different spin channels, main contributions are from $xy$ and $x^2-y^2$ orbitals of vanadium atoms similar to monolayer case. Therefore, the presence of the substrates does not influence significantly on sensitivity of bi- and trilayer VSe$_2$ systems to structural changes. 
 %%%%%%%%%%%%%%%%%%%%%%%%%%%%%%%%%%%%%%%%%%%%%%%%%%%%%%%%%%%%%%%%%%%%%%%%%
%%%%%%%%%%%%%%%%%%%%%%%%%%%%%%%%%%%%%%%%%%%%%%%%%%%%%%%%%%%%%%%%%%%%%%%%%
\subsection{Interlayer binding energy}
To understand the effect of interlayer interactions on structural properties we have
checked interlayer distances and binding energies. The binding energies E$_b$ for
different VSe$_2$ structures were calculated by using the following expression E$_b=($E$-n*$E$_{mono})/m$, where E is the total energy of considered system, E$_{mono}$ - total energy of monolayer, $n$ - number of layers in the considered system, $m$ - average number of interlayer interactions ($m$ = 2, 3/2 and 1 for bulk, 3 and 2-layers,
respectively). Results of these calculations are presented in Table
\ref{tab1}.

In the case when van der Waals interaction is neglected we obtain that the distance between V-V atoms belonging to the same layer is 3.33 \AA \ and the Se-Se interlayer distance equals to 3.12 \AA. When the van der Waals interaction is taken into account such distances equal to 3.31 \AA \ and 3.04 \AA, respectively. In 2- and 3-layer cases we considered the structures (Fig.\ref{fig8}) with the lowest total energies. Calculated values of the binding energies evidence that few-layer VSe$_2$ is pure van der Waals structure in contrast to bulk VSe$_2$ where London dispersion forces is a small addition to electrostatic interactions between V-cations and Se-anions from different
layers. The changes of interlayer distances are proportional to contribution of the dispersion forces to the binding energies (about 0.1 \AA \, in bulk and 0.4 - 0.6 \AA \, in few-layer systems). Therefore, the energy difference in migration barriers in bulk and few-layer VSe$_2$
can be explained by contribution from electrostatic repulsion of anions from the layer
above.
%\small Это какой-то текст
\begin{footnotesize} 
\begin{table}[h!]
\begin{tabular}{c c c c  }
\hline
VSe$_2$  & E$_b$  with & E$_b$  without & Interlayer distance \\
structure & vdW, meV&  vdW, meV &with vdW \\
 & & &(without vdW), \AA\\
\hline
 T-bulk& 7.93&4.79&3.04(3.12)\\
 H-bulk&99.67&95.78&3.22(3.32)\\
 T-two&15.51&-9.36&3.11(3.51)\\
 H-two&24.68&-29.86&3.69(4.27)\\
 T-three&19.05&-90.86&3.08(3.57)\\
 H-three&100.82&-51.77&3.66(4.14)\\
\hline
\end{tabular}
\caption{Interlayer binding energies (meV/formula unit) and interlayer distances calculated for different VSe$_2$ structures with and without vdW interaction.}
\label{tab1}
\end{table}
\end{footnotesize}
 %%%%%%%%%%%%%%%%%%%%%%%%%%%%%%%%%%%%%%%%%%%%%%%%%%%%%%%%%%%%%%%%%%%%%%%%%
%%%%%%%%%%%%%%%%%%%%%%%%%%%%%%%%%%%%%%%%%%%%%%%%%%%%%%%%%%%%%%%%%%%%%%%%%
\subsection{Phonon dispersion}
To complete the picture of physical properties of VSe$_2$ monolayer we have performed calculations of phonon dispersions by using VASP and Phonopy packages \cite{phonopy}. These
combination of the packages is widely used for studying of vibrational properties in
similar systems.\cite{Ersan} For such calculations we used $3\times3\times1$ supercell to obtain sets of forces and mesh grids: $10\times10\times1$ for monolayer and $6\times6\times6$ for bulk. Both H and T phases in nonmagnetic and ferromagnetic configurations were considered. 

\begin{figure}[ht]
    \centering
    \includegraphics[width=1\columnwidth]{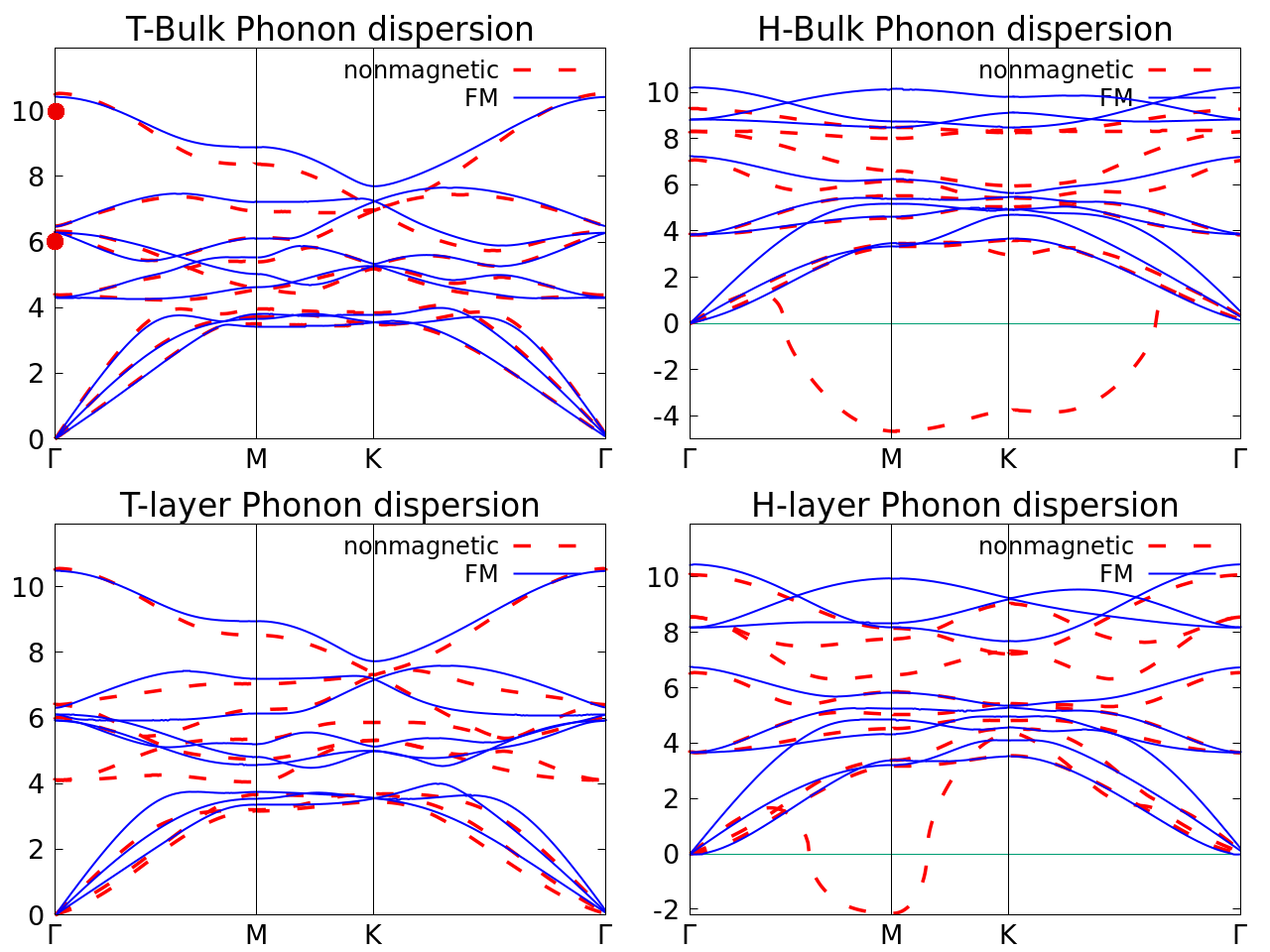}
    \caption{Phonon dispersions calculated for nonmagnetic  (red dashed line) and ferromagnetic state (blue solid line) of monolayer and bulk VSe$_2$. Both T and H phase structures are presented. Red dots denote experimental frequencies taken from Ref.\cite{phonon1}.}
    \label{fig10}
\end{figure}
The calculated phonon spectra are presented in Fig. \ref{fig10}. For the T phase systems (bulk and monolayer) the resulting dispersions demonstrate a weak sensitivity to the magnetism. It is not the case for the H phase configurations. In the nonmagnetic state for H-monolayer and H-bulk we observe a soft phonon mode in the direction $\Gamma-$M$-$K for monolayer and in all symmetry directions for bulk. Existence of such a mode indicates structural instability. Importantly, in the ferromagnetic case the soft mode disappears, which means that the account of magnetism provides structural stability of H phase in both monolayer and bulk. The cause of this effect of magnetic configurations is the robustness of
magnetic interactions (see Fig.\ref{fig6} and discussion above) which is
the same order of magnitude as difference between structural
phase.  For H-bulk and T-monolayer ferromagnetic systems the calculations reveal the appearance of the indirect gap of ~0.57 THz. Comparison of the calculated dispersion curves with available experimental data from Ref.\cite{phonon1} obtained by a point-contact spectroscopy and Raman methods can be fulfilled only for the $\Gamma$ point for which experimental oscillation frequencies are ~6.04 (25 meV) and ~9.67 (40 meV) THz. Our theoretical values of 6.28 and 10.42 THz are in good agreement with experimental data.

%%%%%%%%%%%%%%%%%%%%%%%%%%%%%%%%%%%%%%%%%%%%%%%%%%%%%%%%%%%%%%%%%%%%%%%%%
%%%%%%%%%%%%%%%%%%%%%%%%%%%%%%%%%%%%%%%%%%%%%%%%%%%%%%%%%%%%%%%%%%%%%%%%%
\subsection{Structural phase transition by stretching}
\begin{figure}[ht]
    \centering
    \includegraphics[width=1\columnwidth]{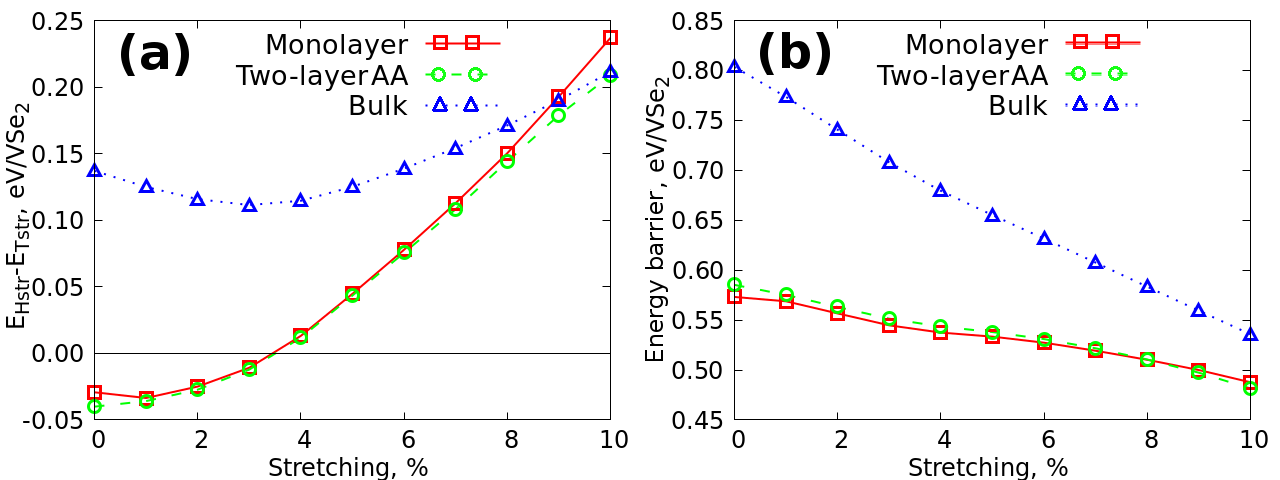}
    \caption{Energy difference between H and T structures of VSe$_2$ (a) and energy barrier (b) as functions of stretching in $a$ and $b$ lattice directions. Lines of different colors correspond to systems with different numbers of layers.}
    \label{fig11}
\end{figure}

The last step of our survey is the modeling of stretch which can appear in the monolayer due to substrate influence. To simulate  this effect, we increase $a$ and $b$ lattice vectors of our structure and then relax atomic positions to find a new ground state corresponding to new lattice parameters. Results of the calculation evidence that a stretching more than 3 percent leads to phase transition of the ground state configuration from H to T in monolayer and bilayer VSe$_2$ (Fig. \ref{fig11}a).

Therefore, the experimentally observed structure\cite{xmcd} of T type can result from a substrate-induced strain. Another effect of the stretching is a decreasing energy barrier for migration between different configurations (Fig. \ref{fig11}b). Here we define the energy barrier as energy difference between T structure and intermediate 30$^{\circ}$ structure 

%%%%%%%%%%%%%%%%%%%%%%%%%%%%%%%%%%%%%%%%%%%%%%%%%%%%%%%%%%%%%%%%%%%%%%%%%%%%%%%%%%%%%%%%%%%%%%
%%%%%%%%%%%%%%%%%%%%%%%%%%%%%%%%%%%%%%%%%%%%%%%%%%%%%%%%%%%%%%%%%%%%%%%%%%%%%%%%%%%%%%%%%%%%%%
%%%%%%%%%%%%%%%%%%%%%%%%%%%%%%%%%%%%%%%%%%%%%%%%%%%%%%%%%%%%%%%%%%%%%%%%%%%%%%%%%%%%%%%%%%%%%%
\section{CONCLUSIONS}
Results of first-principles calculations demonstrate that the energy barrier for the transition between two structural states of VSe$_2$ monolayer with a step-by-step rotation of the single Se atom is rather high. From the other hand the energy cost of the rotation of whole selenium layer is rather low (about 0.60 eV for monolayer and 0.80 eV for bulk). In the case of the monolayer it could be realized with a heating of the samples. The excitation energies of the rotation of the selenium layer up to 10$^\circ$ are very low, therefore, the realistic theoretical description of VSe$_2$ (from monolayer to bulk) should take into account these small deviations from ideal crystal structure. 

Our calculations demonstrate that the transition from the experimentally observed T configuration to the H configuration is accompanied by a considerable change in electronic structure which is a redistribution of $3d$ electrons of vanadium between orbitals. Such transitions significantly influence on transport and thermal properties of VSe$_2$. From the other hand, the values of magnetic moments and total energies of ferro- and antiferromagnetic configurations change gradually between two structural phases. 

In all the considered cases (bulk, few-layer and monolayer) system demonstrate strong favorability of ferromagnetic structure. The analysis of the calculated phonon dispersions has demonstrated a principal role of the ferromagnetism in stabilization of the atomic structure of the VSe$_2$ monolayer in H phase and similar systems. On the basis of the obtained results we can conclude that the experimentally observed paramagnetism in bulk VSe$_2$ and contradictory results of magnetic measurements for monolayers on different substrates are not caused by structural changes. 

The calculations for bi- and trilayers demonstrate that the energy barrier of transition is similar to monolayer. The strain, possibly induced by the substrate, provides the change of the most energetically favorable structure from H to T. Therefore, the experimental observation of T configuration can result from a VSe$_2$ structure stretching by more than 3 percent on substrates. Another effect of the stretching is a decrease of the energy barrier of transition between structural phases. Thus both strain and deviation from ideal structure should be taken into account for realistic description of VSe$_2$ monolayer on substrates. 

\section*{Conflicts of interest}
There are no conflicts to declare.

\section*{Acknowledgements}
This work was supported by the Russian Science Foundation, Grant No. 18-12-00185.

%%%END OF MAIN TEXT%%%


\balance


%%%REFERENCES%%%
\bibliography{ms} %You need to replace "rsc" on this line with the name of your .bib file
\bibliographystyle{rsc} %the RSC's .bst file
\newpage
%%%SI TITLE, AUTHORS AND ABSTRACT%%%
\twocolumn[
  \begin{@twocolumnfalse}
\sffamily
\vspace{2cm}
 \noindent\LARGE{\textbf{Supplementary Information: Structural phase transitions in VSe$_2$: energetics, electronic structure and magnetism$^\dag$}} \\%Article title goes here instead of the text "This is the title"
\vspace{2cm}
 \noindent\large{Georgy V. Pushkarev,\textit{$^{a}$} Vladimir G. Mazurenko,\textit{$^{a}$} Vladimir V. Mazurenko\textit{$^{a}$} and Danil W. Boukhvalov $^{\ast}$\textit{$^{a,b}$}} \\%Author names go here instead of "Full name", etc.

 \end{@twocolumnfalse} \vspace{0.5cm}

  ]
%%%END OF TITLE, AUTHORS AND ABSTRACT%%%

%%%FONT SETUP - please do not change any commands within this section
\renewcommand*\rmdefault{bch}\normalfont\upshape
\rmfamily
\section*{}
\vspace{-1cm}
%FIGURE S vvvv
\renewcommand{\thefigure}{S\arabic{figure}}
%Table S vvvv
\renewcommand{\thetable}{S\arabic{table}}

%%%MAIN TEXT%%%%

\setcounter{figure}{0}
$ $\\
Fig. \ref{figS1} shows angle dependencies of the total energy and magnetic moment in the case of the rotation of the whole Se upper layer of VSe$_2$ monolayer (a,b) and bulk (c,d) within the plane scheme.
\begin{figure}[h]
    \centering
    \includegraphics[width=0.98\columnwidth]{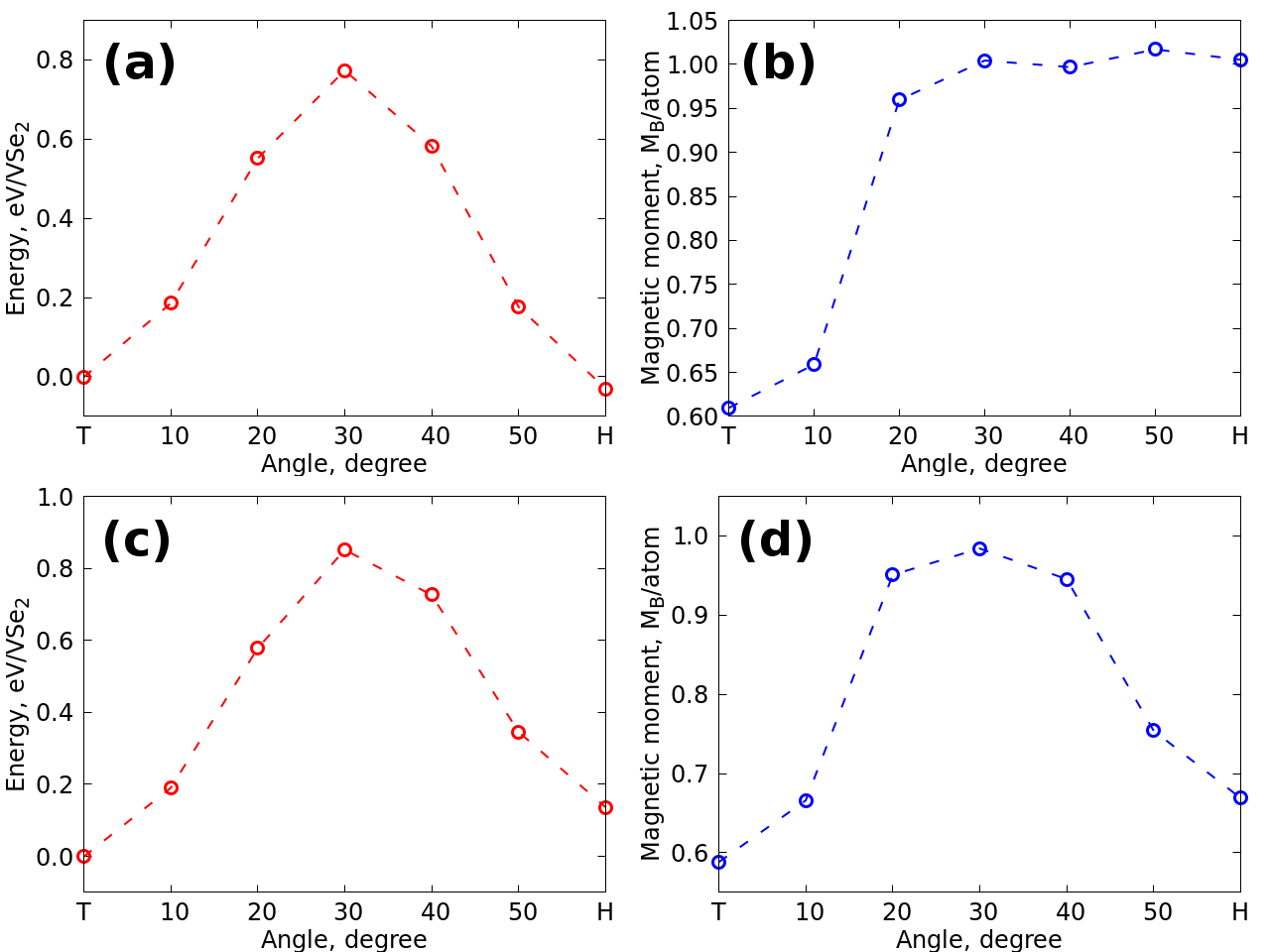}
     \caption{Total energy (a,c) and magnetic moment (b,d) as functions of the rotation angles obtained within the plane model for all Se atoms belonging to the upper layer. The simulations were performed for VSe$_2$ monolayer (a,b) and bulk (c,d).}
     \label{figS1}
\end{figure}

$ $\\ \\ \\ \\ \\ \\ \\ \\

For monolayer one can see that such a rotation scheme is less profitable in energy, since the barrier grows. In turn, the magnetic moment demonstrates the same behavior as with the arc rotation model. In the bulk case we obtain almost the same dependencies, but the maximum of the energy barrier at 30$^{\circ}$ becomes larger than that in the case of the monolayer.

\begin{figure}[h]
    \includegraphics[width=1\columnwidth]{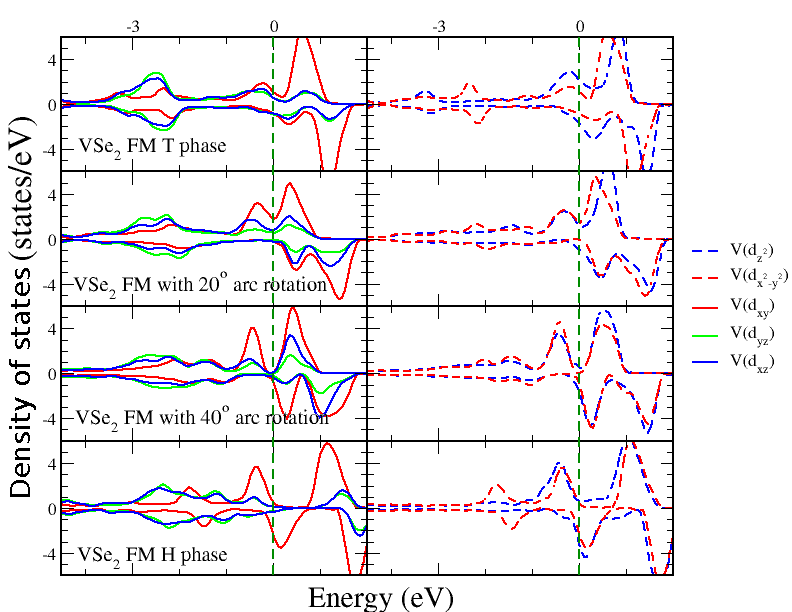}
    \caption{Partial densities of states calculated for VSe$_2$ bulk in the ferromagnetic configuration. The arc rotation scheme  with the $20^{\circ}$ step was used. Left and right panels correspond to ($d_{xy}$, $d_{yz}$, $d_{xz}$) and ($d_{x^2-y^2}$, $d_{z^2}$) sets of states, respectively.}
    \label{figS2}
\end{figure}

 Also we calculated partial densities of states of VSe$_2$ bulk in intermediate points of arc type of rotation with the 20$^\circ$ step (Fig \ref{figS2}).

Figures \ref{figS3} and Fig. \ref{figS4} give band structures of VSe$_2$ monolayer and bulk obtained within arc scheme rotation with the 10$^\circ$ elementary step.

\newpage
\setcounter{figure}{3}
\begin{figure}[ht]
    \centering
    \includegraphics[width=1\columnwidth]{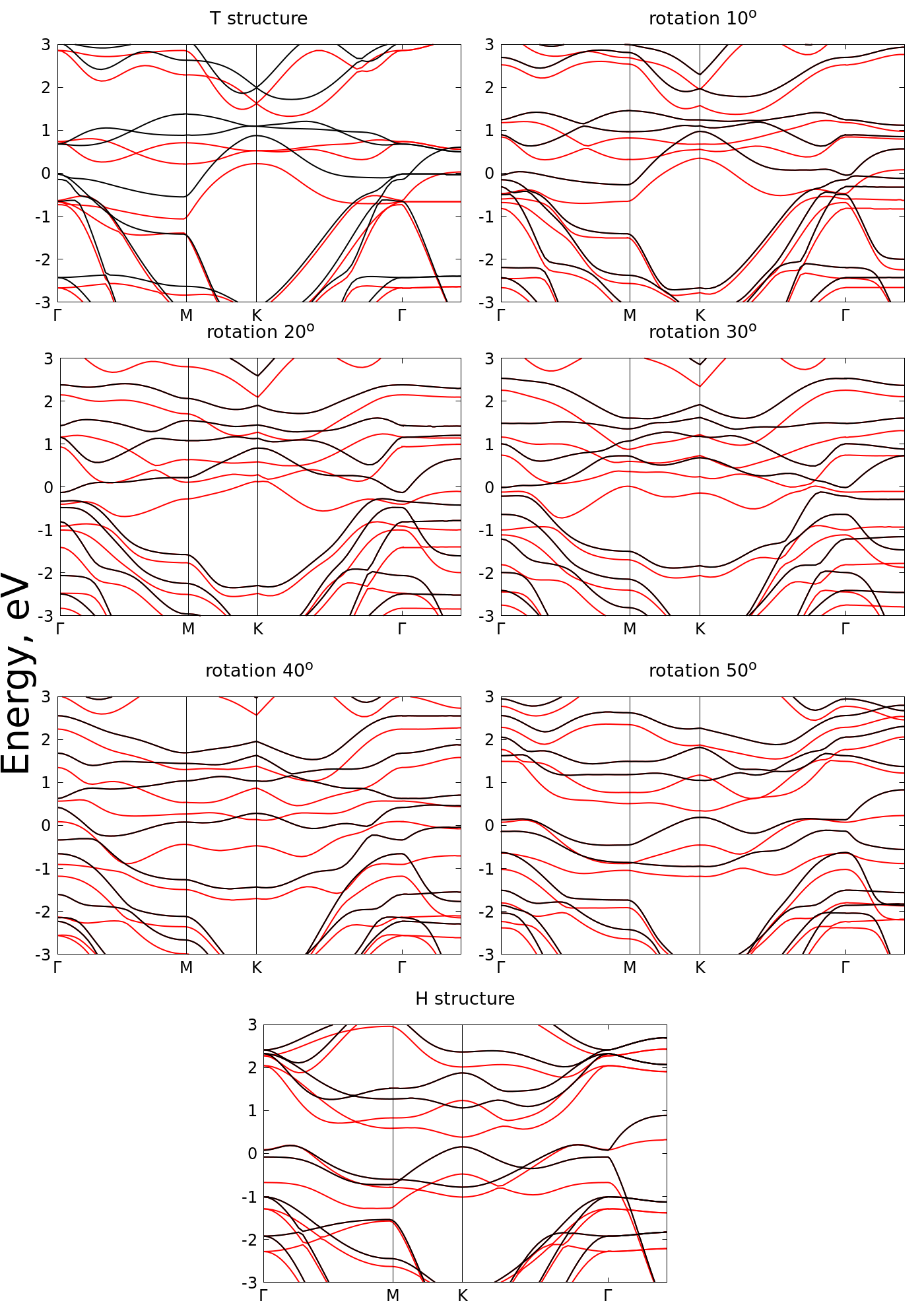}
     \caption{Band structures of the bulk VSe$_2$ crystal calculated for atomic structures modified within the arc rotation model from T phase (0$^{\circ}$) to H phase(60$^{\circ}$) with the step of 10$^{\circ}$. All the calculations were performed for  ferromagnetic configuration. Red lines correspond to spin up states and black ones to spin down.}
     \label{figS4}
\end{figure}
\setcounter{figure}{2}
\begin{figure}[h]
    \centering
    \includegraphics[width=1\columnwidth]{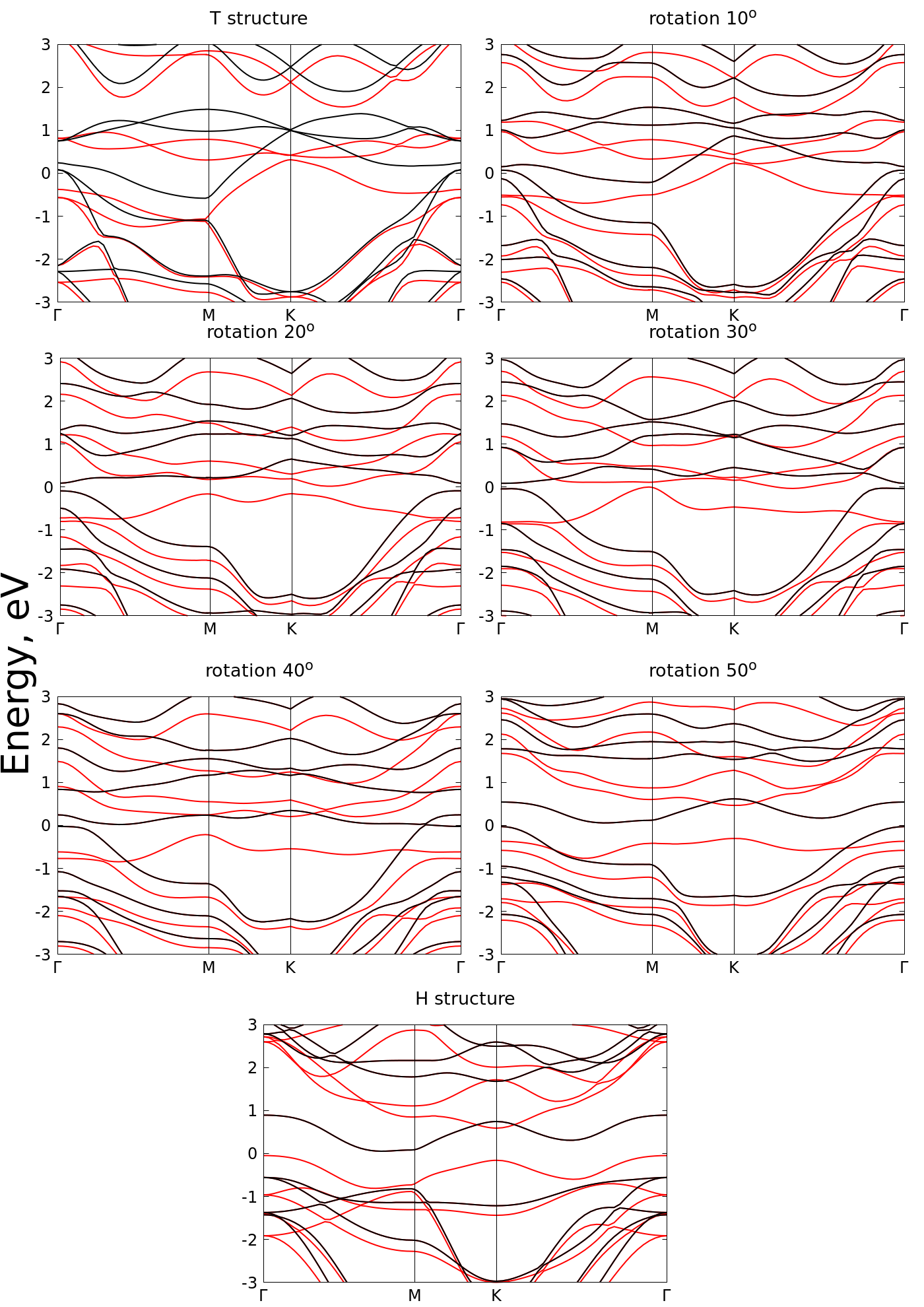}
     \caption{Band structures of the monolayer VSe$_2$ calculated for atomic structures modified within the arc rotation model from T phase (0$^{\circ}$) to H phase(60$^{\circ}$) with the step of 10$^{\circ}$. All the calculations were performed for  ferromagnetic configuration. Red lines correspond to spin up states and black ones to spin down.}
     \label{figS3}
\end{figure}

\end{document}